\documentclass[superscriptaddress,twocolumn,
amssymb,amsmath,nobibnotes,aps,pre,showkeys,showpacs]{revtex4-1}
  
\usepackage{bm}
\usepackage{latexsym}
\usepackage{dcolumn}
\usepackage{amsfonts,amssymb,amsmath}
\usepackage{graphicx,epsfig}
\usepackage{psfrag}
\usepackage{subfigure}
\usepackage{color}
\usepackage{ulem}

\begin{document}

\title{Thermal Properties of a Particle Confined to  a 
Parabolic Quantum Well in 2D Space with Conical Disclination}

\author{Tridev Mishra}\email{tridev.mishra@pilani.bits-pilani.ac.in}
\affiliation{Department of Physics, Birla Institute of Technology and
  Science, Pilani 333031, India.}  \author{Tapomoy Guha
  Sarkar}\email{tapomoy1@gmail.com} \affiliation{Department of
  Physics, Birla Institute of Technology and Science, Pilani 333031,
  India.}  \author{Jayendra
  N. Bandyopadhyay}\email{jnbandyo@gmail.com} \affiliation{Department
  of Physics, Birla Institute of Technology and Science, Pilani
  333031, India.}
\begin{abstract}
The thermal properties of a system, comprising of a spinless
non-interacting charged particle in the presence of a constant external
magnetic field and confined in a parabolic quantum well are
studied. The focus has been on the effects of a topological defect, of
the form of conical disclination, with regard to the thermodynamic 
 properties of the
system. We have obtained the modifications to the traditional
Landau-Fock-Darwin spectrum in the presence of conical
disclination. The effect of the conical kink on the degeneracy
structure of the energy levels is investigated.  The canonical
formalism is used to compute various thermodynamic variables.  The
study shows an interplay between magnetic field, temperature and the
degree of conicity by setting  two scales for temperature
corresponding to the frequency of the confining potential and the
cyclotron frequency of external magnetic field.  The kink
parameter is found to affect the quantitative behaviour of the
thermodynamic quantities. It plays a crucial role in the competition
between the external magnetic field and temperature in fixing the
values of the thermal response functions. This study provides an
important motivation for studying similar systems, however with
non trivial interactions in the presence of topological defects.

\end{abstract}

\maketitle

\section{Introduction}
\label{sec:intro}
In recent decades, advances in nanotechnology, semiconductor device
fabrication and micro-fabrication techniques have thrown open the rich
field of two dimensional electron systems (2DES)
\cite{ando,QDotrev.2002, QDotrev.2002, AshooriNature1996}.  There is a
special focus on systems with confinement along all three spatial
dimensions \cite{reed,Cibert1986,InSbQdot,QDot-review}.  Several
experiments aimed at understanding the electronic \cite{QDotrev.2002}
and optical properties \cite{kash1986, Temkin1987} of such systems,
commonly referred to as quantum dot \cite{reed,makarovsky}, have been
undertaken \cite{smith,ashoori1992,InSbQdot}.  In some of these
experiments the quantum dots are exposed to magnetic fields of varying
strength and their response is studied in terms of electron transport
and inter-band tunneling properties \cite{ashoori1992, InSbQdot, hansen, demel}.

A reasonable model to describe such non-relativistic quantum dot
systems requires a parabolic quantum well as the confining potential
\cite{GaAsQdot}.  However, theoretical exploration of such models is
far from exhaustive and presents several potential situations for
study. One such aspect is the response of a charged particle confined
to a quantum well in an applied magnetic field \cite{tapasheeint} and
constrained to a surface with non-trivial geometry.  An often studied
topological defect is a conical disclination
\cite{mermin79,kroner,Kle,Osipov,furtado94}, which has been the focus
of quantum mechanical problems in curved space
\cite{furtado94,conebrazil,furtado99,furtado01} of the Landau level
type \cite{landau1}.  Another dimension of investigation
\cite{GaAsQdot,jishad} looks into the thermodynamic properties of
confined systems of the Landau-Fock-Darwin \cite{darwin} type in
ordinary Euclidean space.  These lines of examination can be brought
to converge on the issue of thermodynamic behavior of single electron
confined in the presence of a conical disclination, a situation which
has the potency to reveal the physics of quantum dots with novel
geometry. The presence of topological defects in the constraining
surface is expected to affect the thermodynamic characteristics of
such a system and their asymptotic behavior.

 In this paper we analyze the properties exhibited by a charged
 particle constrained on a surface with a defect of the nature of
 conical disclination.  The system comprises of the particle subjected
 to a magnetic field, while it is trapped in a parabolic Fock-Darwin
 potential. The approach is, to first calculate the energy spectrum
 and then use the canonical partition function to uncover the
 thermodynamic properties of the system. We have used the
 Schr\"odinger equation to obtain the energy eigenspectrum.  This
   is motivated by the fact that spectroscopic studies of electronic
   states of quantum dots (such as InSb quantum dot) indicate that a
   Schr\"odinger Hamiltonian with a Fock-Darwin confining potential
   gives reasonable agreement with experiments \cite{InSbQdot}.  We
   introduce a conical disclination defect in such systems through the
   Volterra process \cite{Kle} (discussed in the next section). The
   approach here borrows an idea from gravity, whereby the defect
   appears as a modification of geometry of the underlying space. We
   also note that a similar approach maybe adopted for graphene
   \cite{graphene} like 2D systems. However, the spectrum there is
   linear at the band minima and thereby the Dirac Hamiltonian has to
   be adopted.  We have studied the variations of the thermodynamic
 quantities of interest like internal energy, specific heat and
 entropy with magnetic field, temperature and {\it extent of the
   defect}. The asymptotic limits of these are checked for
 confirmation with expected results.

 The paper consists of four sections. Sec. \ref{sec:formalism} is
 dedicated to developing the mathematical formalism.  The defect is
 introduced as a modification of the metric from its otherwise
 Euclidean form.  Beginning with a suitable choice of coordinates, the
 Hamiltonian of the system is constructed. The Schr\"odinger equation
 is then solved for this Hamiltonian to obtain the energy
 spectrum. This is followed by obtaining the various thermodynamic
 variables of the system using the canonical partition function.  The
 expressions for these quantities are recast in terms of dimensionless
 parameters and their behavior is studied. The asymptotics are checked
 for consistency. In Sec. \ref{sec:results} we present the results of
 our study. Finally we conclude with a discussion and summary in the
 last section.

\section{Formalism}
\label{sec:formalism}
The topological defect being introduced in the current study is a
conical disclination. This entails a two-dimensional (2D) conical
space which is locally flat at all points except for the origin
\cite{conebrazil}.  The construction of this space is to be visualized
as the consequence of cutting out a sector with a certain apex angle
called the {\it deficit angle}, from the ordinary 2D flat space and
subsequently welding together the newly revealed edges
\cite{volterra}. The metric for such a space, in the usual polar
coordinates $(r,  \phi)$ is given by $ g_{\mu \nu} = {\rm diag}  ( 1, ~ r^2)$.
However, it has to be kept in mind that $\phi$ here has an incomplete
angular range $[0, \, 2\pi\kappa]$ with $ \kappa \neq 1$.  This, being
a consequence of the surgical procedure performed previously. The
parameter $\kappa$ is a measure of the deficit angle. It quantifies
the conicity of the surface and shall henceforth be referred to as the
{\it kink parameter}. The kink here represents a singular deformity of
the 2D conical surface at the origin.  The metric described above can
be expressed in terms of the complete angular coordinate $\theta$ as
follows
\begin{eqnarray}
\label{metric2}
 ds^{2}=  \kappa^{-2}d{\rho}^{2} + {\rho}^{2}d\theta^{2}.
\end{eqnarray}
where, $\theta$ varies in $[0, \, 2\pi]$.  The transformation from
plane polar coordinates to the new coordinate system, i.e. from $
\left( r,\phi \right ) \rightarrow \left( \rho,\theta \right) $ is
achieved via the set of transformation equations
\begin{equation}
\rho=\kappa{r} ~~  ~~ \theta=\kappa^{-1}\phi.
 \end{equation} 
The curvature is measured by the quantity 
\[
2\pi \frac{ \kappa - 1 }{\kappa} \delta^{(2)} (\rho), \] where
$\delta^{(2)}(\rho)$ is the Dirac delta function in two dimensions
\cite{furtado01}.  Hence, for $ 0 < \kappa < 1 $ we have negative
curvature and for $ 1 < \kappa < \infty $ the curvature at origin is
positive. We note that the  metric described here in the context of 
2D condensed matter system  also  arises in the description
of space-time around a cosmic string \cite{vilenkin}.

 In the above described space we consider a charged spin-less quantum
 particle ( for our purposes it has electronic mass and charge). This
 particle is subjected to a constant magnetic field $\mathbf{B}$ which
 is normal to the conical surface. The appropriate choice of magnetic
 vector potential that yields such a magnetic field is given in the
 symmetric gauge by
 \begin{eqnarray}
  \mathbf{A}(\rho)=\frac{B\rho}{2\kappa}\hat{e}_\theta
\end{eqnarray}
where $ B = | {\mathbf B}|$.
This gives rise to the standard  quantized single particle Landau level states \cite{landau1}.

In order to model the confinement of the particle within a small
region on the surface, we subject the particle to a parabolic potential
of the Fock-Darwin type \cite{darwin} given by

\begin{eqnarray}
 V(\rho)=\frac{1}{2}M\omega_{p}^{2}\frac{{\rho}^2}{{\kappa}^2}
\end{eqnarray}
where $M$ is the effective mass of the particle and $\omega_{p}$ is a measure of
the steepness of the confinement. The appearance of the kink parameter
indicates that the background space is conical.  The choice of such a
potential is motivated by symmetry considerations and its frequent
appearance in the modeling of quantum dots with low occupancy \cite{refer-tapash}.

The Hamiltonian for the particle of mass $M$,  assumed to be carrying  a negative charge of magnitude $e$ under 
minimal electromagnetic coupling,  is given in the cone space coordinates $\left ( \rho, \theta \right)$ as  
\begin{eqnarray}
\begin{split}
 H&=-\frac{{\hbar}^{2}}{2M}\left[\frac{{\kappa}^{2}}{\rho}\frac{\partial}{\partial\rho}\left(\rho\frac{\partial}{\partial\rho}\right)+
   \frac{1}{{\rho}^{2}}\frac{\partial^2}{{\partial}\theta^{2}}\right]
 -
 \frac{i}{2}\frac{\hbar\omega_{c}}{\kappa^2}\frac{\partial}{\partial\theta}\\ &
 \quad + \frac{1}{8}M\omega_{c}^{2}\frac{{\rho}^{2}}{{\kappa}^{4}} +
 \frac{1}{2}M\omega_{p}^{2}\frac{{\rho}^2}{{\kappa}^2}
\end{split}
\label{Hamiltonian}
 \end{eqnarray}
where the parameter $\omega_{c}$
introduced here is the cyclotron frequency $ \omega_{c}= eB/Mc$.
Note the appearance of the kink parameter $\kappa \neq 1$ when one expresses
the Hamiltonian in the cone space.

The general form of eigenfunctions for this Hamiltonian can be
guessed from symmetry arguments.  Separation of the Schr\"odinger
equation into radial and angular components yields such a general
form 
\begin{eqnarray}
 \psi(\rho, \theta) = \frac{1}{\sqrt{2\pi}}{e}^{im\theta} R_{nm}(\rho)
\end{eqnarray}
The quantum numbers $n$ and $m$ are to be defined using the
appropriate boundary conditions.  Here, $R_{nm}(\rho)$ stands for the
radial component of the wave function. The condition on $m$ is readily
obtained by requiring $\psi$ to be unique under a rotation of $2\pi$,
ie $\psi(\rho, \theta) =\psi(\rho, \theta + 2 \pi)$. This implies that
$m$ has to be an integer. The Schr\"odinger equation $H
\psi_{nm}=E_{nm}\psi_{nm}$ yields the following equation for the
radial wave function $R_{nm}(\rho)$.
\begin{eqnarray}
\begin{split}
 &-\frac{{\hbar}^{2}}{2M}\left[\frac{{\kappa}^{2}}{\rho}\frac{\partial}{\partial\rho}\left(\rho\frac{\partial}{\partial\rho}R_{nm}(\rho)\right)-\frac{{m}^{2}}{{\rho}^{2}}R_{nm}(\rho)\right]\\ &
  +
  \left(\frac{1}{2}\frac{\hbar\omega_{c} m }{\kappa^2} +\frac{1}{8}M\omega_{c}^{2}\frac{{\rho}^{2}}{{\kappa}^{4}}
  +
  \frac{1}{2}M\omega_{p}^{2}\frac{{\rho}^{2}}{{\kappa}^{2}}\right)R_{nm}(\rho)\\ &
  =E_{nm}R_{nm}(\rho)
 \end{split}
\label{eq:rad}
 \end{eqnarray}
The procedure to solve the above equation is through a set of standard
transformations,  which involves the  introduction of a new parameter $\Omega$
with dimension of frequency. The parameter $\Omega$ is  given by 
\begin{eqnarray} \Omega=\sqrt{\omega_{p}^{2} +
  {\left(\frac{\omega_{c}}{2\kappa}\right)}^{2}}. 
\label{omega}
\end{eqnarray}
 Following the formalism in \cite{furtado94} eq. \eqref{eq:rad} can be transformed to a 
form which permits solution in terms  of the confluent-hypergeometric
function (see  Appendix). Our primary interest lies in the energy levels which are given by  
 \begin{eqnarray}\label{EE}
  E_{nm}=\left(2n + 1 + \frac{|m|}{\kappa}\right)\hbar\Omega + \frac{m\hbar\omega_{c}}{2{\kappa}^{2}}
 \end{eqnarray}

If we consider the system to be at equilibrium with a heat bath at
temperature $T$, the canonical partition function shall be given by
 \begin{eqnarray}
 {\mathcal Z}={\displaystyle\sum_{n,m}e^{-\beta(2n + 1)\hbar\Omega} \,
   e^{-\beta\left[\frac{|m|}{\kappa}\hbar\Omega +
       \frac{m\hbar\omega_{c}}{2{\kappa}^{2}}\right]}}.
\label{partition1}
\end{eqnarray}
 where $\beta= \frac{1}{k_{B}T}$ and $ k_{B}$ is the Boltzmann
 constant.  The sum is over the discrete energy levels given in
 Eq. \eqref{EE}. 
Introducing dimensionless variables $\chi_1=
\frac{\beta\hbar\Omega}{\kappa}$ and $\chi_2=
\frac{\beta\hbar\omega_{c}}{2{\kappa}^{2}}$ the above expression maybe
simplified to
\begin{eqnarray}
 \mathcal{Z}= \frac{\sinh \chi_1}{4\sinh\left(\dfrac{\chi_1 +\chi_2}{2}\right)\sinh\left(\dfrac{\chi_1-\chi_2}{2}\right)\sinh(\kappa \chi_1)}.~~~~~~~
\label{partition2}
\end{eqnarray}
\newline{}
It is now possible to compute thermodynamic quantities from this expression
of the partition function.

The internal energy $U$ for the system is given by
\begin{eqnarray}
 &&U = -\frac{\partial\ln{\mathcal{Z}}}{\partial\beta} \nonumber \\
&&= -\biggl\{\chi_1\coth(\beta \chi_1) - \frac{\chi_1 + \chi_2}{2}\coth\beta\left(\frac{\chi_1+\chi_2}{2}\right)\nonumber\\ 
  &-&\frac{\chi_1 - \chi_2}{2}\coth\beta\left(\frac{\chi_1-\chi_2}{2}\right)-\chi_1\kappa\coth\beta\kappa \chi_1\biggr\}
\end{eqnarray}
Similarly one can obtain the specific heat capacity $C_{\rm v}$
\begin{eqnarray}
  &C_{\rm v} &=
  k_{B}{\beta}^{2}\frac{{\partial}^{2}\ln{\mathcal{Z}}}{\partial{\beta}^{2}}
  \nonumber \\ &=& k_{B}{\beta}^{2}\biggl\{ \frac{{\left(
      \chi_1+\chi_2\right)}^{2}}{4}{\rm
    csch}^{2}\beta\frac{\left(\chi_1+\chi_2\right)}{2}+
  \chi_1^{2}{\kappa}^{2}{\rm csch}^{2}\beta\kappa \chi_1 \nonumber
  \\ &+&\frac{{\left( \chi_1-\chi_2\right)}^2}{4}{\rm
    csch}^{2}\beta\frac{\left(\chi_1-\chi_2\right)}{2}-{\chi_1}^{2}{{\rm
      csch}}^{2}\beta \chi_1 \biggr\}.
\end{eqnarray}
The Helmholtz free energy $ F = - \ln \mathcal{Z}/\beta  $ may be
used to calculate the entropy $S$ as $S=(U-F)/ T$. This yields
the following expression
\begin{eqnarray}
\begin{split}
 &S = \frac{1}{T}\biggl\{ -\chi_1\coth(\beta \chi_1) + \frac{\chi_1 +
    \chi_2}{2}\coth\left(\beta
  \,\frac{\chi_1+\chi_2}{2}\right)\\ &+\frac{\chi_1 -
    \chi_2}{2}\coth\left(\beta\frac{\chi_1-\chi_2}{2}\right)+\chi_1\kappa\coth(\beta\kappa
  \chi_1)\biggr\}\\ & + k_{B}\biggl\{ \ln \sinh(\beta \chi_1) - \ln
  \sinh\left(\beta \frac{\chi_1+\chi_2}{2}\right) \\ &-\ln
  \sinh\left(\beta \frac{\chi_1-\chi_2}{2}\right) - \ln
  \sinh(\beta\kappa \chi_1) - \ln4 \biggr\}
 \end{split}
\end{eqnarray}
We shall now study the variation of these quantities with the external
magnetic field $B$ and temperature $T$. In order to facilitate this,
it is helpful to choose certain special units which render the
physical quantities $U$, $C_{v}$ and $S$ dimensionless.  We introduce
a parameter $\alpha = \omega_{c} / \omega_{p}$ to quantify the
magnetic field strength in units of $M \omega_p c/ e$ and $ \xi = k_B
T / \hbar \omega_p$ to represent temperature measured in units of $
\hbar \omega_p/ k_B$. We also introduce $ \widetilde{\alpha} =
\sqrt{1+ \alpha^2/4\kappa^2}$ and $\alpha_{\pm} = \widetilde{\alpha} \pm \alpha/2 \kappa$. 

Using these new dimensionless
parameters, we have the internal energy $U$, entropy $S$ and specific
heat $C_V$ may be expressed as

\vspace{0.2cm}

{\it Internal  energy} :
\begin{eqnarray}
\begin{split}
  &\frac{U}{\hbar \omega_p} = \frac{1}{\kappa}\biggl\{
  -\widetilde{\alpha}\coth\left(\frac{\widetilde{\alpha}}{\xi\kappa}\right)
  + \frac{\alpha_{+}}{2}\coth \left(
  \frac{\alpha_{+}}{2\xi\kappa}\right) \\ & +
  \frac{\alpha_{-}}{2}\coth \left(\frac{\alpha_{-}}{2\xi\kappa}\right)
  +
  \widetilde{\alpha}\kappa\coth\left(\frac{\widetilde{\alpha}}{\xi}\right)\biggr\}
 \end{split}
 \end{eqnarray}

{\it Specific heat :}
\begin{eqnarray}
  \begin{split}
 &\frac{C_{v}}{k_B} = \frac{\widetilde{\alpha}^2}{\xi^2}{\rm
      cosech^2}\left(\frac{\widetilde{\alpha}}{\xi}\right) +
    \frac{{\alpha_{+}}^2}{4\kappa^2\xi^2} {\rm
      cosech}^2\left(\frac{\alpha_{+}}{2\kappa \xi}\right)\\ &
    +\frac{{\alpha_{-}}^2}{4 \kappa^2 \xi^2} {\rm
      cosech}^2\left(\frac{\alpha_{-}}{2\kappa \xi}\right)
    -\frac{\widetilde{\alpha}^2}{\xi^2\kappa^2}{\rm
      cosech^2}\left(\frac{\widetilde{\alpha}}{\kappa\xi}\right)
 \end{split}
\end{eqnarray}

{\it Entropy :}
\begin{eqnarray}
 \begin{split}
  &\frac{S}{k_B}
   =\frac{1}{\xi\kappa}\biggl\{-\widetilde{\alpha}\coth\left(\frac{\widetilde{\alpha}}{\xi\kappa}\right)
   + \frac{\alpha_{+}}{2}
   \coth\left(\frac{\alpha_{+}}{2\xi\kappa}\right) \\ & +
   \frac{\alpha_{-}}{2}\coth
   \left(\frac{\alpha_{-}}{2\xi\kappa}\right) +
   \widetilde{\alpha}\kappa\coth\left(\frac{\widetilde{\alpha}}{\xi}\right)\biggr\}\\ &
   +\biggl\{\ln\sinh\left(\frac{\widetilde{\alpha}}{\xi\kappa}\right)
   - \ln\sinh\left(\frac{\alpha_{+}}{2\xi\kappa}\right) & \\ &
   -\ln\sinh\left(\frac{\alpha_{-}}{2\xi\kappa}\right)-\ln\sinh\left(\frac{\widetilde{\alpha}}{\xi}\right)-
   \ln4\biggr\}.
 \end{split}
\end{eqnarray}
The asymptotic behaviour of the above expressions in the low
temperature limit is instructive to look at. The internal energy $U$
in the low temperature limit is given by $U \rightarrow \hbar\Omega$, where
 $\Omega$ is defined earlier in eq.\eqref{omega}. The low
temperature asymptotic form $( \xi \rightarrow 0 )$ of entropy $S$ is
given by
\begin{eqnarray}
\label{eq:asym-ent}
\begin{split}
 S & \approx \left(1 + \frac{\alpha_{+}}{\xi \kappa} \right)
 e^{\frac{-\alpha_{+}}{\xi\kappa}} + \left( 1 + \frac{\alpha_{-}}{\xi
   \kappa} \right) e^{\frac{-\alpha_{-}}{\xi\kappa}} \\ & + \left( 1 +
 \frac{2 \widetilde\alpha}{\xi }\right)
 e^{\frac{-2\widetilde\alpha}{\xi}} - \left( 1 + \frac{2
   \widetilde\alpha}{\xi \kappa}\right)
 e^{\frac{-2\widetilde\alpha}{\xi\kappa}}
 \end{split}
\end{eqnarray}
 The specific heat
in the low temperature limit, is approximated by the following function
of temperature.
\begin{eqnarray}
\begin{split}
 C_v &
 \approx\frac{4\widetilde{\alpha}^2}{\xi^2}e^{\frac{-2\widetilde{\alpha}}{\xi}}
 + \frac{\alpha_{+}^2}{\xi^2\kappa^2}e^{\frac{-\alpha_{+}}{\xi\kappa}}
 \\ & +
 \frac{\alpha_{-}^2}{\xi^2\kappa^2}e^{\frac{-\alpha_{-}}{\xi\kappa}} -
 \frac{4\widetilde{\alpha}^2}{\xi^2\kappa^2}e^{\frac{-2\widetilde{\alpha}}{\xi\kappa}}
\end{split}
\end{eqnarray}

\section{Results and Discussion}
\label{sec:results}
\begin{figure}[h]
\includegraphics[height=6.0cm, width=8.5cm]{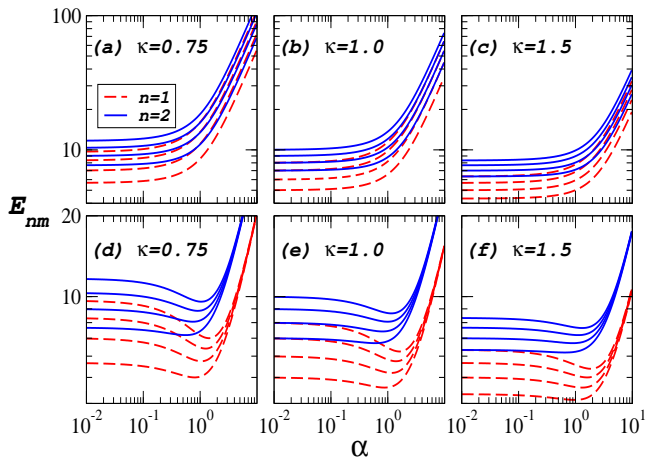}
\caption{(Color online) The low lying energies of the Landau-Fock-Darwin energy
  spectrum for various values of the kink parameter $\kappa$. The
  upper panel (a-c) shows the spectra for positive values of the
  quantum number $m = 2, 3, 4, 5$ (lower to the upper) 
and the lower panel (d-f)
  corresponds to negative values of $m = -2, -3, -4, -5$ with decreasing 
magnitude $|m|$ from upper to the lower curves.}
\label{fig:Evsalpha}
\end{figure}

\subsection{The Energy Spectrum}
The Landau-Fock-Darwin energy spectrum is given by Eq. \eqref{EE}.
Fig. (\ref{fig:Evsalpha}) shows the variation of $E_{nm}$ with the
external magnetic field parameter $\alpha$, for a few chosen values of
the kink parameter $\kappa = 0.75, 1.0, 1.5$. The behaviour of the
energy levels is different for positive and negative values of the
quantum number $m$.  The figure shows the variation of $E_{nm}$ with
$\alpha$ for $n = 1, 2$. In the upper panel we show the case when the
integer $m$ is assumed to take positive values 2, 3, 4 and 5 for each
$n$. The behaviour at very low magnetic field shows that $E_{n m}$ is
independent of $\alpha$ for $\alpha \lesssim 10^{-2}$. In this low
magnetic field regime one finds the usual degeneracies of $(n,m)$
pairs since $ E_{nm} \approx ( 2n + 1 + |m|/\kappa )\hbar
\omega_p$. In our case with $\kappa = 1.0$ this occurs, for example
between $(n, m)$ pairs like $ [(2,2) , (1,4)] $, $[(2,3), (1,5)] $ and
$[(2,4), (1,6)]$.  These degeneracies starts to get lifted when the
external magnetic field is sufficiently high ($ \alpha \approx 1$).
At very high magnetic fields ($ \omega_c >> \omega_p$) and for $m >0$,
we have $ E \rightarrow [(2n + 1 )/2\kappa + m/\kappa^2]\hbar
\omega_c$ leading to new degeneracies.  In the relatively high
magnetic field region of $\alpha\approx 10$ one can readily observe
that curves for all $(n,m)$ are monotonically increasing with nearly
fixed slopes.  The transition between these extreme behaviours occurs
in the intermediate field region of $\alpha\approx 1$.  We note,
  that in the intermediate and large magnetic field regions the
  difference between the energy levels with the same value of $n$ but
  different values of $m$ is larger as compared to the low field
  region. This is owing to the fact that $\omega_c$ is larger for
  higher magnetic fields. For example the level corresponding to $(1,
  3)$ is higher than $(2, 2)$.  The energy levels shift in magnitude
for changing $\kappa$ which implicitly affects the degeneracy pattern.
\noindent

\begin{figure}[b]
\includegraphics[height=6cm, width=8.5cm]{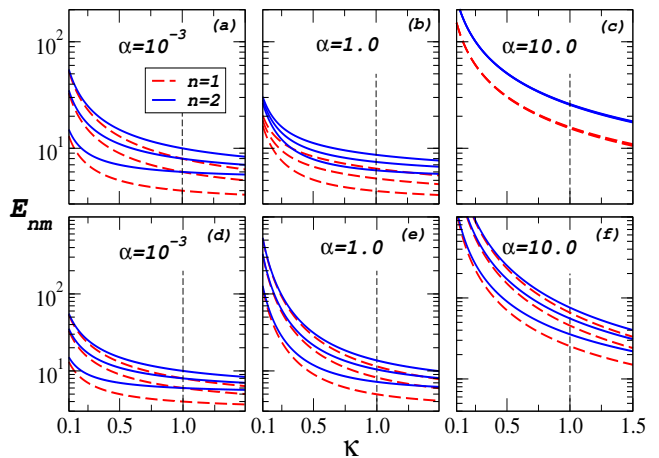}
\caption{ (Color online) The first few levels of the Landau-Fock-Darwin energy
  spectrum as a function of the kink parameter $\kappa$. The
  upper panel corresponds to negative values of the quantum number $m
  = -1, -3, -5$. The lower panel shows the same for positive values of
  $m = 1, 3, 5$. Three magnetic field values are chosen with $\alpha =
  10^{-3}, 1.0, 10.0$.}
\label{fig:Evskappa}
\end{figure}

The figures in the lower panel 1(d)-1(f) shows the spectrum for
negative $m$ values. The low magnetic field behaviour is the same as
for the positive $m$ case. However at large magnetic fields the
  term $ m \hbar \omega_c / 2 \kappa^2 $ starts to play an important
  role and cancels the term $ |m| \hbar \Omega/2 \kappa $ in this
  regime. The spectrum becomes independent of $m$ and only depends on
$n$. The increase of $E_{nm}$ is approximately linear with magnetic
field $\alpha$. The transition between the two regimes occurs again at
$\alpha \approx 1$.

Figure \ref{fig:Evskappa} shows the variation of energy with the kink
parameter $\kappa$ for three different values of the applied magnetic
field. Fig 2(a), (b) and (c) show the variation for negative $m$
values ($ m = -1, -3, -5$) corresponding to $n= 1, 2$. The curves show
a monotonic decrease of $E_{nm}$ with $\kappa$ in all the three
regimes of magnetic field $\alpha$. The value $\kappa = 1$ corresponds
to the case with no topological defect. We note an asymmetry in the
nature of variation of $E_{nm}$ about this value of $\kappa$.  The
energy levels are a decreasing function of $\kappa $ for both $\kappa
\geq 1$ and $\kappa < 1 $ showing that positive and negative deficit
angles point towards fundamentally different physical situations. The
expression for $E_{nm}$ diverges as $ \kappa \rightarrow 0$. This, 
  however is of no real consequence since $\kappa = 0 $ corresponds to
  an unphysical divergent curvature at the origin.

The vertical dotted line indicating the case without any defect
($\kappa = 1$) passes through the point of intersection of the energy
levels.  These points correspond to the degenerate energy levels at
low magnetic field.  The degeneracy of the $(n, m)$ levels for $
\kappa = 1$ are seen to get lifted for $\kappa \neq 1$ as the energy
levels for different $m$ vary differently with $\kappa$.  In Fig. 2(c)
the different $m$ levels for a given $n$ are degenerate and remain so,
irrespective of $\kappa$.  The figures 2(d)-2(f) show a similar
variation for positive $m$ values.  Whereas the degeneracies at weak
magnetic field ( Fig.  2(d)) gets lifted for $\kappa \neq 1$ there are
new degeneracies that are created at higher magnetic fields. This is
seen in Fig. 2(e)-2(f) where non-degenerate energy levels at $\kappa =
1$ intersect each other at $\kappa \neq 1$ showing the  emergence of
accidental degeneracies that did not exist in the defect free theory.

\subsection{Thermodynamic properties}
\begin{figure}[b]
\includegraphics[height=6cm, width=8.5cm]{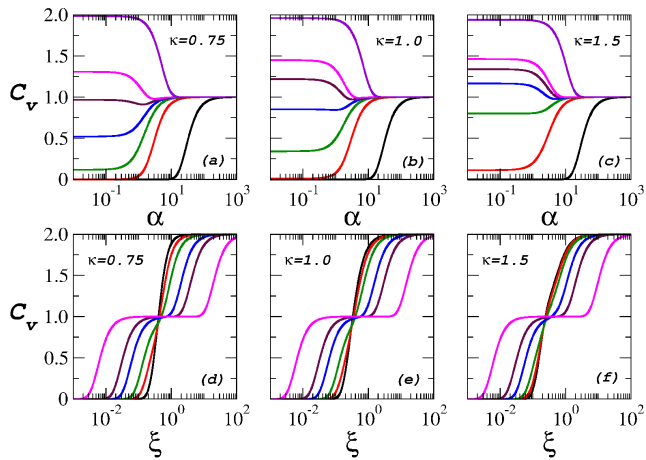}
\caption{ (Color online) The $C_{\rm v}$ shown here is in units of $k_B$. The upper
  panel shows the variation of $C_{\rm v}$ with magnetic field for
  various values of temperature $\xi = 0.01, 0.1, 0.2, 0.3, 0.4, 0.5,
  2.0$ (curves from lower to upper). At high magnetic fields $C_{\rm
    v}$ attains the value $1.0$ as the spectrum reduces to the free
  Landau levels with no confinement. The high temperature value of
  $C_{\rm v}$ for moderate to low magnetic fields is $2$ as the
  confinement term dominates at these regimes. The Lower panel shows
  the variation of $C_{\rm v}$ with temperature for $ \alpha = 0.001,
  1.0, 2.0, 5.0, 10.0, 50.0$ (left to right in the upper right corner
  of the figures). Here again the plateau in $C_{\rm v}$ is seen for
  the high magnetic fields and only at high temperatures $C_{\rm v}$
  attains the value $2.0$. }
\label{fig:fig3}
\end{figure}

\begin{figure*}[t]
\begin{tabular}{lcr}
\includegraphics[height=4.70cm,width= 5.2cm]{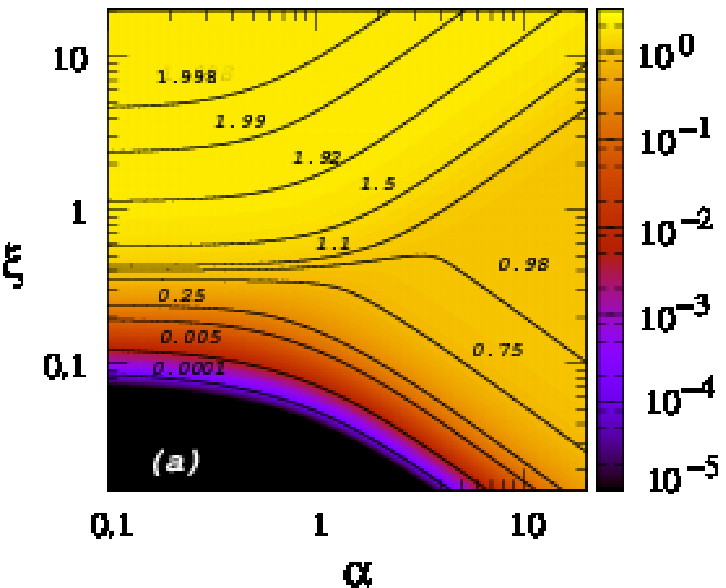}&
\includegraphics[height=4.7cm,width= 5.3cm]{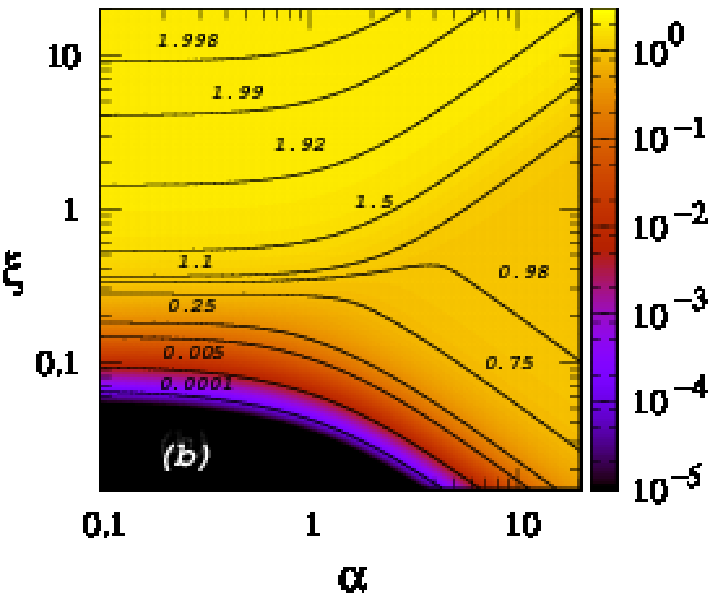}&
\includegraphics[height=4.7cm,width=5.2cm]{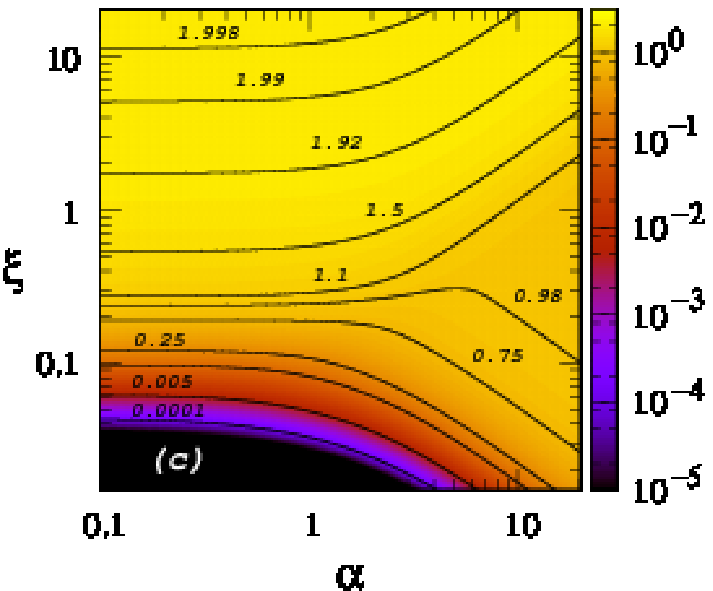}
\end{tabular}
\caption{ (Color online) The contour map for specific heat $c_{\rm v}$ in the $(\xi,
  \alpha)$ phase plane, for three values of the kink parameter (a)
  $\kappa = 0.75$ , (b) $ \kappa = 1.0$, (c) $\kappa = 1.5$. }
\label{fig:fig4}
\end{figure*}

\begin{figure}[b]
\includegraphics[height=6cm, width=8.5cm]{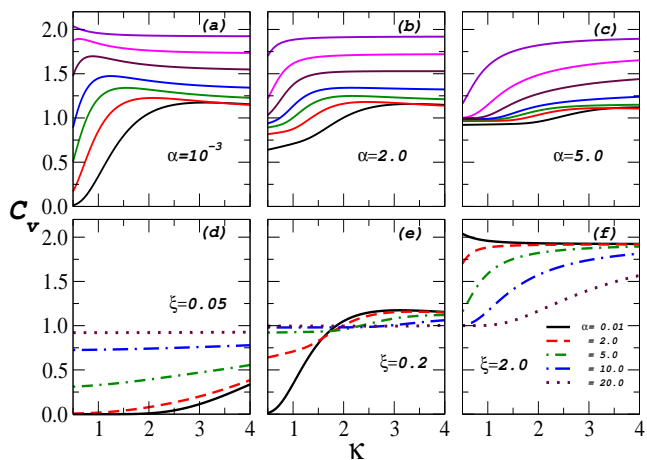}
\caption{ (Color online) The dependence of $C_{\rm v}$ on the kink  parameter $\kappa$ for different values of external magnetic field $\alpha$ and temperature $\xi$. For the upper panel $\xi = 2.0, 1.0, 0.7, 0.5, 0.4, 0.3, 0.2$ (top to bottom).}
\label{fig:fig5}
\end{figure}

The non-interacting spinless charged particles are assumed to be in
equilibrium with a heat reservoir at temperature $T$. The starting
point of the thermodynamic analysis is the evaluation of the partition
function for the energy spectra given in Eq. \eqref{EE}.  The
Landau-Fock-Darwin Hamiltonian has two energy scales associated with
the two frequencies $\omega_p$ (which fixes the strength of the
parabolic confinement) and $\omega_c$, the cyclotron frequency related
to the external magnetic field.  The relative strengths of these
frequencies are expected to govern the equilibrium behaviour of the
system.  The thermodynamic properties of interest, depend on the
temperature $\xi$ and external magnetic field $\alpha$, expressed in
our chosen convenient energy unit $\hbar \omega_p$.  The parameters in
the Hamiltonian $(\omega_p, \omega_c, \kappa)$ have a crucial
interplay in determining the responses of the system.  The $\kappa =
1$ case with no defects has been studied in earlier works
\cite{GaAsQdot, jishad}. It is important to note that for $\kappa = 1$,
the limiting behaviour of the system for $\omega_p \rightarrow 0$ (or
equivalently $\omega_c >> \omega_p$) and $ \omega_c \rightarrow 0$ are
entirely different and describe two completely distinct physical
situations.  The former describes a pure Landau problem of a free
particle without any confinement, whereas the latter describes a
particle in a two dimensional parabolic well without a coupling to an external
magnetic field.  The $\omega_p \rightarrow 0$ limit has a pure
quantum mechanical Landau-level spectra of a one-dimensional
oscillator and has the degeneracy that depends on the size of the
system.  The energy spectra for the case $\omega_c \rightarrow 0$
mimics that of a 2D oscillator.  The translational symmetry of the
pure Landau level situation is lost completely in the other extreme
limit of a pure confinement problem.  The general Landau-Fock-Darwin
solution interpolates between these  extreme cases.  In the
presence of $\kappa \neq 1$ the same qualitative features are
expected. However, the role of $\kappa$ needs to be
explored and is subsequently discussed in this paper.

We follow the Gibbs formalism to compute thermodynamic quantities like
free energy, entropy and specific heat. In this approach, the
thermodynamic response functions are obtained as derivatives of the
partition function.  The canonical partition function (see
Eq. \ref{partition1}) is obtained for the Hamiltonian in
Eq. \eqref{Hamiltonian}. In the final form, this partition function
(see Eq. \eqref{partition2}) is seen to diverge in the limit $\omega_c
>>\omega_p$ (or $\alpha >> 1$) since, $\chi_1 $ and $\chi_2$ are equal in this
limit. This singularity of the partition function, when the
confinement strength is vanishingly small, has been addressed in
\cite{jishad} and maybe regularized by putting certain cutoffs to the
smallest value that $\omega_p$ can take. This cutoff depends on the
temperature and the degeneracy of the pure Landau level. The
thermodynamic quantities like $ F, U, S$ and $C_{\rm v}$ however,
manifest no such singularity.

Figure \ref{fig:fig3} shows the variation of $C_{\rm v}$ with magnetic
field $\alpha$ and temperature $\xi$ for different values of the kink
parameter $\kappa$. The variation of $C_{\rm v}$ with $\alpha$ shows
that for weak external magnetic field and low temperatures $C_{\rm}$
asymptotically approaches {\it zero}. However, in this weak $\alpha$ regime,
at high temperatures $C_v \rightarrow 2k_B$ asymptotically. This is in
consonance with the equipartition principle. The low  $\alpha$ end behaves
like a 2D oscillator (hence the factor 2). In the high magnetic field
regime ($\alpha$ large), $C_{\rm v}$ saturates to $k_B$. This
region corresponds to the pure Landau level with the energy spectrum of an
1D oscillator.  The qualitative features are similar when $\kappa \neq
1$. However, we see that changing $\kappa$ from $0.75$ to $1.5$
continuously, leads to a shift of the curves from the lower end towards
the upper. This can be qualitatively ascribed to the fact that
$\kappa$ appears as a multiplicative factor to $\xi$ in the expression
for $C_{\rm v}$ and a change of $\kappa$ roughly amounts to a recalibration
of the temperature scale.
 
The variation of $C_{\rm v}$ with temperature $\xi$ is shown in the
lower panel of Fig.  \ref{fig:fig3}.  When the value of $\alpha$ is
small, the rise of $C_{\rm v}$ with temperature is steep, and in a
very small temperature range, $C_{\rm v}$ rises from zero to a stable
value of $2k_B$. Beyond the transition temperature, $C_{\rm v}$
remains flat at this value.  In this situation the system is
essentially dominated by the parabolic confining potential and the
physics of the Landau levels is missing.  The situation is
considerably different when $\alpha$ is large. Here the effect of
confinement is weak and $C_{\rm v}$ attains a plateau like level when
temperature is increased. The value of $C_v$ remains constant at $k_B$
for a range of temperatures after which it rises to $2k_B$ only at
high values of $\xi$. The formation of the plateau can be ascribed to
the dominance of the Landau 1D oscillator spectrum at high magnetic
fields as opposed to the 2D oscillator spectrum of the parabolic well
when the magnetic field is weak. The extent of the plateau region is
found to be sensitive to $\kappa$. We shall discuss this $\kappa$
dependence later.

Figure \ref{fig:fig4} shows the contour map of $C_{\rm v}$ in the
$(\alpha, \xi)$ plane. At very low temperatures, $C_{\rm v}
\rightarrow 0$ except, when the external magnetic field is large. The
lower left corner of the $(\alpha, \xi)$ plane corresponds to this
phase where $C_{\rm v}$ is small.  Increasing the temperature at small
values of $\alpha$ leads to a monotonic increase of $C_{\rm v}$ to its
saturated value of $2 k_{B}$ (upper left corner of the phase
diagram). At such low values of $\alpha$ there is hardly any Landau
coupling to the magnetic field.  The Landau plateau occurs at large
$\alpha$ when the energy spectrum approaches the Landau levels. This
is the forked region of the contour map, where, for a considerable
range of intermediate temperatures the value of $C_{\rm v}$ remains at
the $k_B$ level, and only increases to $2 k_B$ at still higher
temperatures (this is not seen in the phase diagram and occurs for
values of $\xi$ even above the upper right corner). The extent of the
forking region (plateau in $C_{\rm v}$ depends on the kink
parameter. Infact, it is seen to decrease with increasing
$\kappa$. This can be understood by noting that a changing $\kappa$
can be equivalently seen as changing $\xi$ with a fixed $\kappa$.  The
qualitative features of the phase diagram remain the same when $\kappa$
is varied. However, there are quantitative changes which we shall
discuss now.

\begin{figure}[b]
\includegraphics[height=6cm, width=8.5cm]{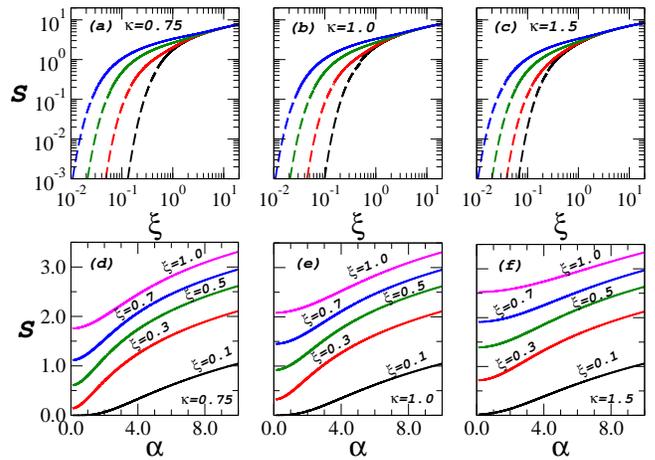}
\caption{(Color online) The upper panel (a-c) shows the variation of $S$ with $\xi$ for various values of $\alpha$ and $\kappa$ with $\alpha = 10.0, 5.0, 2.0, 0.1$ (from left to right). The broken lines
  indicate the temperature range for validity of the low temperature
  asymptotic behaviour of $S$. The lower panel shows the $\alpha$ dependence of 
entropy for specific temperatures and $\kappa$.}
\label{fig:fig6}
\end{figure}

Figure \ref{fig:fig5} shows the variation of $C_{\rm v} $ with
$\kappa$. At high temperatures, $C_{\rm v}$ is not sensitive to
$\kappa$ unless the magnetic field $\alpha$ is also very high. This is
seen in the figures 5(a)-(c). The specific heat is however very
sensitive to $\kappa$ at low temperatures. Increasing $\kappa$ can be
equivalently interpreted as a scaling of $\alpha$ and this explains
the plateau (characteristic of large $\alpha$) when $\kappa$ is
large. At large $\alpha$ (figure 5(c)), all the low temperature curves
cluster around the $k_B$ level and stabilizes at the $2k_B$ level only
for high temperatures.  Figure 5(e) shows that there is a cross over
of $C_{\rm v}$ at a certain value of $\kappa$. This implies that at
some intermediate low temperatures $C_{\rm v}$ is not much sensitive
to the changes in the magnetic field for certain values of
$\kappa$. At higher temperatures, however, $C_{\rm v}$ saturates to
$2k_B$. This growth is slower for the curves corresponding to large
$\alpha$ values which tends to stay in the plateau region as compared
to the case when $\alpha$ is small. Here, we see that $\kappa$
essentially re-calibrates the temperature scale.

Figure \ref{fig:fig6} shows the behaviour of the entropy as a function
of magnetic field and temperature. The competition between the
variables $\xi$ and $\alpha$ decides the degree of order in the
system. We find that the asymptotic form of $S$ in
Eq. \eqref{eq:asym-ent} is valid for a certain value of $\xi$ that
depends on the magnetic field and $\kappa$. This region of validity of
this limiting form of entropy is shown by broken lines in Fig.
\ref{fig:fig6} (a)-(c).  The third law of thermodynamics is respected
and we have $S \rightarrow 0 $ as $\xi \rightarrow 0$.  The growth of
entropy from the low temperature ordered regime to the disordered
state at high temperature, depends on the magnetic field.  The growth
is steeper for higher magnetic fields. However at very high
temperatures the magnetic field dependence keeps decreasing.  Figures
6 (d-f) shows the variation of entropy with magnetic field.  At very
high magnetic field there is a slowing down on the rate at which $S$
increases. This feature is seen for a wide range of temperatures.  The
effect of $\kappa$ here is clearly that of a scaling parameter that
re-calibrates the temperature scale $\xi$.

\section{Summary and Conclusion}
\label{sec:summary}

In this work, we have carried out a study of the thermodynamic
ramifications of a conical defect, in the context of Landau-Fock-Darwin problem. The competing behaviour of the temperature and
magnetic field is noted, and how a change in
the kink parameter influences this. The variation of quantities like specific heat and entropy with the kink parameter illustrates the physical effect of the disclination to
be a sort of recalibration of the temperature scale. Also of note are
the essential non trivialities inherent in the Landau-Fock-Darwin
problem with respect to the symmetry of the system. These are
recovered here in the presence of the conical defect as is illustrated by the
step in the specific heat curve  at high magnetic fields, which reflects the
interpolation of the behaviour between a 1D and a 2D oscillator.

We  conclude by noting that it is possible to extend this analysis to further studies which could incorporate  discrete lattice structure and interactions
in the presence of this class of topological defects.

\appendix*
\section{}
To solve the radial eigenvalue equation \eqref{eq:rad},  we introduce 
$\zeta = \rho^2 M \Omega/\hbar$. This transformation yields the following equation
\begin{eqnarray}
\zeta \frac{\partial^2 R(\zeta)}{\partial \zeta^2}  + \frac{\partial R(\zeta)}{\partial \zeta}  + \Xi (\zeta)   R(\zeta) = 0  
\end{eqnarray}
where we have used
\[
 \Xi(\zeta) = \frac{\beta}{\kappa^2} - \frac{\zeta}{4 \kappa^4} -\frac{m^2}{4\kappa^2 \zeta} 
~~~~{\rm and}~~~~~~
\beta = \frac{1}{2} \left( \frac{E_{mn}}{\hbar \Omega} - \frac{\omega_c m}{2 \kappa^2 \Omega} \right).
\]
Using variables $\zeta' = \zeta/ \kappa^2$ and $m' = m/\kappa$ we have
\begin{eqnarray}
\zeta'\frac{\partial^2 R(\zeta')}{\partial \zeta'^2}  + \frac{\partial R(\zeta')}{\partial \zeta'}  + \Xi' (\zeta')   R(\zeta') = 0  
\label{a2}
\end{eqnarray}
where the new function $\Xi'$ is 
\[
 \Xi'(\zeta') = \beta' - \frac{\zeta'}{4} -\frac{m'^2}{4\zeta'}, ~~~~{\rm with}~~~~
\beta' = \frac{1}{2} \left( \frac{E_{mn}}{\hbar \Omega} - \frac{\omega_c m'}{2 \kappa\Omega} \right).
\]
Assuming $R(\zeta')$ to be of the form 
\[ R(\zeta') = e^{-\frac{\zeta'}{2}} \zeta'^{\frac{|m'|}{2}} Y(\zeta'), \]
the equation \eqref{a2} reduces to
\begin{eqnarray}
\zeta' \frac{\partial^2 Y}{\partial \zeta'^2} &+& \left( |m'| + 1 - \zeta' \right ) \frac{\partial Y}{\partial \zeta'} \\ \nonumber
&+& \left ( \beta' - \frac{|m'|}{2}  - \frac{1}{2} \right) Y = 0.
\end{eqnarray}
The solution to this equation is given in terms of the confluent-hypergeometric function as
\begin{equation}
Y(\zeta') = F \left [ - \left( \beta' - \frac{|m'|}{2} -\frac{1}{2} \right),~ |m'| +1 ;~  \zeta' \right]
\end{equation}

The requirement of boundedness of $R(\zeta')$ as $\zeta' \rightarrow \infty$ is met if 
\begin{equation}
 \beta' - \frac{|m'|}{2} -\frac{1}{2} = n,
\end{equation}
where $n$ is a non-negative integer. From this boundary condition (after substituting $m/\kappa$  in place of $m'$) the eigenenergies are given by 
\begin{eqnarray}
  E_{nm}=\left(2n + 1 + \frac{|m|}{\kappa}\right)\hbar\Omega + \frac{m\hbar\omega_{c}}{2{\kappa}^{2}}
 \end{eqnarray}
The eigenfunctions corresponding to these eigenvalues are obtained after imposing the requirement that for integral values of $n$, the confluent hypergeometric function reduces to Laguerre polynomials given as
\begin{equation}
L^{\alpha}_n(\zeta')= \frac{\Gamma\left(\alpha+n+1\right)}{\Gamma\left(\alpha+1\right)n!}F\left(-n,~\alpha+1;~\zeta'\right)
\end{equation}
here $\Gamma(n)=(n-1)!$ is the usual  gamma function.Thus the eigenfunctions are of the form
\begin{eqnarray}
R(\zeta)=Ce^{-\frac{\zeta}{2\kappa^2}}{\left(\frac{\zeta}{\kappa^2}\right)}^{\frac{|m|}{2\kappa}}L^{\frac{|m|}{\kappa}}_n(\zeta)
\end{eqnarray}
where $C$ is the constant of normalization.The first term in the product represents a Gaussian in the variable $ \rho$ whose spread is now determined by the degree of disclination. The localization of the wave function is hence sensitive to  $\kappa$ and consequently all probability densities are affected by the degree of conicity.The appearance of $|m|/\kappa$ indicates the deficit/surplus of the polar angle quantified through $ \kappa$. 

\end{document}